\documentclass[preprint,12pt]{elsarticle}

\usepackage{amssymb}

\newcommand{\qbb}         {{$Q_{\beta\beta}$}}
\newcommand{\thalfzero}   {${T^{0\nu}_{1/2}}$}

\newcommand{\onbb}        {{$0\nu\beta\beta$}}
\newcommand{\nnbb}        {{$2\nu\beta\beta$}}
\newcommand{\twonu}       {{$2\nu\beta\beta$}}
\newcommand{\gerda}       {\textsc{Gerda}}
\newcommand{\majorana}    {\textsc{Majorana}}
\newcommand{\gess}        {{$^{76}$Ge}}
\newcommand{\geenr}       {{$^{\rm enr}$Ge}}          

\newcommand{\kgyr}        {{kg$\cdot$yr}}
\newcommand{\IGEX}        {\textsc{Igex}}
\newcommand{\hdm}         {\textsc{HdM}}
\newcommand{\ctsper}      {cts/(keV$\cdot$kg$\cdot$yr)}

\usepackage{lineno}

\journal{Physics of the Dark Universe}

\begin{document}

\begin{frontmatter}



\title{Status of double beta decay experiments using isotopes other than 
$^{136}$Xe}


\author[lngs,lns]{L.~Pandola}

\address[lngs]{INFN, Laboratori Nazionali del Gran Sasso, Assergi (AQ), Italy}
\address[lns]{INFN, Laboratori Nazionali del Sud, Catania, Italy}
\ead{pandola@lngs.infn.it}

\begin{abstract}
Neutrinoless double beta decay is a lepton-number violating process 
predicted by many extensions of the standard model. It is actively searched 
for in several candidate isotopes within many experimental projects. 
The status of the experimental initiatives which are looking for the 
neutrinoless double beta decay in isotopes other than $^{136}$Xe is 
reviewed, with special emphasis given to the projects that 
passed the R\&D phase. \\
The results recently released by the experiment \gerda\ are also summarized 
and discussed. The \gerda\ data give no positive indication of 
neutrinoless double beta decay of \gess\ and disfavor in a model-independent 
way the long-standing observation claim on the same isotope. The lower
limit reported by \gerda\ for the half-life of neutrinoless double beta decay 
of \gess\ is \thalfzero$> 2.1\cdot 10^{25}$~yr (90~\% C.L.), or 
\thalfzero$ > 3.0\cdot 10^{25}$~yr, when combined with the results 
of other \gess\ predecessor experiments.
\end{abstract}


\end{frontmatter}


\section{Neutrinoless double beta decay}\label{sec:intro}
Many extensions of the standard model of particle physics predict the 
existence of the neutrinoless double beta (\onbb) decay: 
\begin{displaymath}
(A,Z) \to (A,Z+2) + 2 e^-.
\end{displaymath}
Such a transition violates by two units 
the lepton number conservation and it is thus forbidden by the standard 
model. The \onbb\ decay is actively searched for in different 
candidate isotopes by many experimental programs throughout the world:  its 
observation would bring far-reaching consequences and demonstrate that 
neutrinos are Majorana 
particles~\cite{bil12,ver12,rodejohann,gomez12,sche82}. In the assumption 
that the decay is mainly driven by the exchange of light Majorana 
neutrinos, it is possible to establish an absolute scale for the 
neutrino mass, provided that the nuclear matrix elements are known. 
The experimental signature of \onbb\ is a mono-energetic peak at the 
$Q$-value of the decay (\qbb).\\
A claim of observation of the \onbb\ decay of $^{76}$Ge was made 
more than ten years ago~\cite{klapdor2001}, based on the 
re-analysis of the data of the Heidelberg-Moscow experiment~\cite{hdm}. The 
net \onbb\ signal reported in Ref.~\cite{klapdor1} is 
$(28.75 \pm 6.86)$~events. 
The same Ref.~\cite{klapdor1} 
reports the most probable value of the half-life of the decay, 
\thalfzero=$1.19 \cdot 10^{25}$~yr: this indirectly provides the 
proportionality factor which links the number of counts to the inverse 
of the half-life, and which depends on experimental parameters, like  
efficiency and exposure. Being the number of events proportional to 
1/\thalfzero, the one-sigma range for \thalfzero\ can be calculated as 
\thalfzero$=(1.19^{+0.37}_{-0.23}) \cdot 10^{25}$~yr. Later, the same 
group re-analyzed the data and strengthened the claim~\cite{klapdor2}, 
by using a novel event selection method based on the pulse shape 
information\footnote{The number of events ascribed to the \onbb\ 
decay is $11.32 \pm 1.75$~events, which is converted to 
\thalfzero$=(2.23^{+0.44}_{-0.31}) \cdot 10^{25}$~yr in 
Ref.~\cite{klapdor2}.}.
However, major inconsistencies  in the calculation of the half-life 
\thalfzero\ from the number of events presented in Ref.~\cite{klapdor2}, 
which should involve an efficiency factor of the pulse shape selection, 
were pointed out recently~\cite{bernhard}. \\ 

This paper summarizes the status of the experiments worldwide which are 
looking (or will look in the future) for the \onbb\ decay in isotopes other 
than $^{136}$Xe $-$ that are described in a separate paper~\cite{piepke}. 
A special focus is given to the projects that passed 
the R\&D phase and are now in a more advanced development stage 
(design, construction, commissioning, or data taking) and to the 
experiment \gerda, which recently reported results on \onbb\ decay of \gess. 

\section{Overview of the present projects}
\begin{table}[htbp]
\centering \caption{\label{tab:experiments} Compilation of the present active  
initiatives searching for \onbb\ in isotopes other than $^{136}$Xe. For each of them, it is 
reported the candidate isotope, the location and the development status.}
\begin{tabular}{llll} 
Project & Isotope & Location & Status \\
\hline 
\gerda~\cite{gerda:2013:tec,gerda:2013:prl} & $^{76}$Ge & LNGS & Phase~I completed \\
      &           &      & Upgrade to Phase~II \\
Cuore-0/Cuore~\cite{cuore2004,cuore2014} & $^{130}$Te & LNGS & Data taking \\
 & & & Commissioning \\
\majorana\ Demonstrator~\cite{majo2014} & $^{76}$Ge & SURF & Commissioning \\
SuperNEMO Demonstrator~\cite{snemo} & $^{82}$Se & LSM & R\&D, Construction \\
SNO+~\cite{snoplus2011} & $^{130}$Te & SNOLAB & R\&D, Construction \\
CANDLES~\cite{candles} & $^{48}$Ca & Kamioka & R\&D, Construction \\
Cobra~\cite{cobra2005} & $^{116}$Cd & LNGS & R\&D \\
Lucifer~\cite{lucifer2013} & $^{82}$Se & LNGS & R\&D \\
DCBA~\cite{dcba2010} & many & [Japan] & R\&D \\
AMoRE~\cite{amore2012} & $^{100}$Mo & [Korea] & R\&D \\
MOON~\cite{moon2010} & $^{100}$Mo & [Japan] & R\&D \\
\hline
\end{tabular} 
\end{table}
One of the main goals of all neutrinoless double beta decay experiments in 
the last decade was to confirm or refute the observation 
claim. Due to the large uncertainties in the calculations of nuclear matrix 
elements for the \onbb\ candidate isotopes (e.g. Ref.~\cite{bup13} for a 
compilation of recent calculations in \gess\ and $^{136}$Xe), experiments 
employing \gess\ are best suited to scrutinize the observation 
claim, by the direct comparison of \thalfzero. 
On the other hand, the observation of the \onbb\ in a different 
isotope would be mandatory to confirm and certify a discovery. \\ 

Many candidate \onbb\ isotopes are being considered in experimental 
programs or R\&D, notably $^{76}$Ge, $^{130}$Te, $^{136}$Xe, $^{82}$Se, 
$^{150}$Nd, $^{100}$Mo, $^{48}$Ca and $^{116}$Cd. These nuclei are 
characterized by their own \qbb-values and nuclear matrix 
elements (NME), 
which make them more or less favorable in terms of decay rate for a given 
neutrino effective mass 
and of expected background.
However, it has been 
shown~\cite{bernhard,robertson2013,bar13} that the phase space and the NME 
effects nearly compensate, so that the specific decay rates are the same 
within a factor of two for all candidate isotopes above.
In particular, an effective neutrino mass of a few 
tens of meV would yield a decay rate of approximately 
1~decay/(ton$\cdot$yr) in all isotopes. \\
Given the lack of a ``golden'' candidate isotope for \onbb\ searches, the 
choice is mostly driven by practical or experimental grounds, such as:  
easiness of isotope enrichment, energy resolution, half-life of 
the neutrino-accompanied double beta decay (\twonu), scalability and 
modularity of the design, and of course cost. A somewhat special role 
is played by \gess, because of the historical reasons and because of the 
observation claim made with it.\\ 
Exposures in the scale of tens of \kgyr\ have been achieved so far 
in \gess, $^{136}$Xe, $^{130}$Te and $^{100}$Mo. The experiments under 
development and construction are hence aiming to the scale of the hundreds of 
\kgyr, which allows for the initial assay of the ``inverted hierarchy'' 
region~\cite{vissani2002}.

Table~\ref{tab:experiments} reports an inventory of the present active 
\onbb\ projects using isotopes other than $^{136}$Xe. 
They are in different phases 
of development, ranging from data taking to initial R\&D. The projects that 
are in the construction phase or data taking are briefly reviewed in 
the following. 

\subsection{Cuore-0 and Cuore}
The Cuore experiment~\cite{cuore2004} at the INFN Gran Sasso Laboratory 
(LNGS), Italy, 
is going to search for the \onbb\ of $^{130}$Te (\qbb=2528~keV) by using the 
bolometric technique~\cite{fiorini84} which was already demonstrated by 
MiBETA~\cite{mibeta} and Cuoricino~\cite{cuoricino2008}. The 
Cuoricino precursor experiment collected an exposure of 19.7~\kgyr\ and gave 
a lower limit on the half-life of $^{130}$Te, 
\thalfzero$> 2.8 \cdot 10^{24}$~yr~\cite{cuoricino2011}. \\
The Cuore experiment will deploy 988 crystals of TeO$_2$ (741~kg in total, 
206~kg of $^{130}$Te), 
arranged in a large array made of 19 towers, each containing 52 crystals in 
13 planes. The crystals will be operated in a cryostat, at a temperature of 
a few mK. The target background at \qbb\ is 0.01~cts/(keV $\cdot$ kg 
of TeO$_2$ $\cdot$ yr). \\
The first step of the 
Cuore assembly is Cuore-0: it is a full Cuore tower, which has been 
cooled down and operated in the old Cuoricino cryostat since March 2013. 
Cuore-0 served to define and test the Cuore assembly procedures, but it is also a 
real data-taking experiment, having 51 crystals out of 52 properly working.
It demonstrated that a complete Cuore tower can be assembled in less 
than four weeks. Calibrations performed with a $^{232}$Th source show that the 
spectroscopic performances and the energy resolution (5.7~keV FWHM at \qbb) 
are well within expectations~\cite{cuore2014}. The Cuore-0 background measured  
from the initial 7.1~\kgyr\ exposure is 0.07~\ctsper, limited by the radioactivity 
of the old Cuoricino cryostat~\cite{cuore2014}. Given the background and the energy 
resolution, Cuore-0 is expected to surpass with one year of live time the 
$^{130}$Te \onbb\ half-life sensitivity achieved by Cuoricino.\\
Meanwhile, the construction of the full Cuore experiment is regularly ongoing, 
concerning both the assembly of the detector towers (which is foreseen to be 
completed in July, 2014) and the commissioning of the cryogenic plant. The 
first cool down of the set up is expected by end of 2014. The sensitivity 
of the full-scale experiment after 5~yr of data taking and with a 
background of 0.01~\ctsper\ is about $10^{26}$~yr 
(90\% C.L.)~\cite{cuoresensitivity}.

\subsection{Majorana Demonstrator}
The \majorana\ demonstrator (MJD) project plans to construct and operate large modular 
arrays of high-purity germanium (HPGe) detectors for the search of the \onbb\ decay 
of \gess~\cite{majo2014}. The set up is located underground at the 4850-feet level 
of the SURF Laboratory, United States. 
The main goal of the project is to demonstrate that an appropriate 
low-background level can be achieved to justify the design and the 
construction of a \gess\ ton-scale experiment. \\
The target background in a 4-keV wide region of interest (ROI) at the \qbb\ of 
the \gess\ decay (\qbb=2039~keV) is 3~counts/(ROI$\cdot$ton$\cdot$yr), after all 
analysis cuts. This scales to 1 count/(ROI$\cdot$ton$\cdot$yr) in a one-ton experiment. 
The MJD set up is designed to be very compact: the array of HPGe detectors is hosted in 
two independent ultra-clean vacuum cryostats, made out of electroformed copper. The 
cryostats are surrounded by a low-background passive shielding made of copper and lead, 
with an active muon veto. In the first phase, 40~kg of HPGe p-type point-contact 
detectors will be deployed (20~kg in each cryostat), 30~kg of which are made out 
of germanium isotopically enriched in \gess\ (\geenr). \\
The first step of the commissioning was performed in 2013 with a prototype vacuum 
cryostat, having the same design as the final cryostats but made out of non-electroformed 
copper. Two strings of natural detectors were deployed and operated. 
The subsequent steps of the MJD commissioning will be the
operation of the first cryostat (seven strings of \geenr\ detectors with a few natural
detectors), in Summer 2014, and of the second cryostat (three strings of \geenr\ 
detectors and four strings of natural detectors), in about Summer 2015.
\subsection{SuperNEMO Demonstrator}
The SuperNEMO project~\cite{snemo} is the successor of the completed 
NEMO3 experiment~\cite{nemo3}, which 
was in operation at the Modane underground Laboratory (LSM), France, between 2003 and 2011. 
The main design feature of the experiment is the tracking capability, allowing 
to detect separately the two electrons emitted in the \onbb\ decay. 
The source is in the form of very thin foils and does not coincide with 
the detector. This gives the maximum flexibility in the choice of the candidate \onbb\ 
isotopes: actually seven of them were studied in NEMO3, the most important one 
being $^{100}$Mo. \\
The NEMO3 detector was composed
by 6180 drift chambers (Geiger cells) for the tracking part, and 1940 plastic scintillators 
(read out by photo-multipliers) for the calorimetry part. 
A key experimental figure of merit is the energy resolution 
(8\% FWHM at \qbb), which determines the background due to the \nnbb\ decay. The 
background level achieved in NEMO3 
at the \qbb=3034~keV of $^{100}$Mo was $1.2 \cdot 10^{-3}$~\ctsper, 
mainly ascribed to \nnbb\ decay and to $^{222}$Rn-induced events: all other background 
sources are efficiently suppressed by the topological reconstruction. No events were 
found in the high-energy range [3.2-10.0] MeV in a total exposure of 47~\kgyr\ (collected 
with many isotopes). 
Recently new results have been released by NEMO3 on the \onbb\ 
decay of $^{100}$Mo: 18 events 
were observed in the range [2.8-3.2] MeV, to be compared with $16.4 \pm 1.3$ expected from 
background, after a 34.7~\kgyr\ exposure. This turns out in a lower limit on the half-life
\thalfzero $> 1.1 \cdot 10^{24}$~yr (90\% C.L.) \cite{snemo13}. \\
The SuperNEMO project~\cite{snemo} will use the same design and technology 
which were successfully employed in NEMO3. The goal is to deploy up to 100 kg of 
target isotope within 20 identical modules at LSM. $^{82}$Se is the primary 
choice as the target isotope but $^{150}$Nd and $^{48}$Ca are also considered, depending 
on the development of viable enrichment procedures. The SuperNEMO Collaboration will firstly 
operate one module (7~kg of $^{82}$Se) as a demonstrator: it is presently under construction 
at LSM. Each module will contain 2000 drift chambers for the tracking part and 712 plastic 
scintillators for the calorimetry; it will be shielded by iron (300 tons) and water. 
In order to meet the background specifications, the 
energy resolution was improved by a factor of two in SuperNEMO, namely to 4\% (FWHM) at \qbb. 
Furthermore, very stringent limits must be achieved for the $^{208}$Tl, $^{214}$Bi and $^{222}$Rn 
radioactivity of the source foils. In this case, it is anticipated that the SuperNEMO demonstrator 
can run background-free for 7~kg of $^{82}$Se and two years of data taking. 
\subsection{SNO+}
The facilities and infrastructures of the former SNO heavy-water neutrino experiment~\cite{sno}
at SNOLAB, Canada, are being refurbished and upgraded to support a new project, 
named SNO+. In particular, heavy water is replaced by liquid scintillator as the 
main target, thus providing a much superior light yield. Given the low background 
and the tracking capability, SNO+ has several physical reaches 
(e.g. supernova and solar neutrinos), but priority has been granted to the \onbb\ 
decay searches. The candidate \onbb\ isotope is $^{130}$Te, which will be deployed in the 
detector in the form of 0.3\% loading of the liquid 
scintillator~\cite{snoplus2011,billersnoplus:2013}. The total mass of 
$^{130}$Te in the fiducial volume (3.5~m) would hence be 800 kg. The main experimental 
issue is to achieve a sufficient energy resolution (i.e. a sufficient light  
yield from the loaded scintillator) such to suppress the background from 
the two-neutrino decay. For an energy resolution $\sigma$=4\% at \qbb=2528~keV, a 
potential sensitivity is expected to 200 meV neutrino effective mass in two years 
of data taking. \\
The milestones anticipated by the SNO+ Collaboration are to fill the SNO detector with 
purified liquid scintillator in the mid of 2014, and then to load the scintillator with 
Te-based compounds in early 2015.
\subsection{CANDLES}
CANDLES is a project which aims to look for the \onbb\ decay of $^{48}$Ca by using 
CaF$_2$ detectors~\cite{candles,candles2}. 
The decay has a very high \qbb-value (4272~keV), which makes the experiment 
practically insensitive to the environmental  $\gamma$ background. Given the very 
low natural abundance of $^{48}$Ca (0.187\%), isotopic enrichment is mandatory for 
a competitive experiment: R\&D is ongoing to identify 
and optimize a viable enrichment strategy. The basic design of the project is to operate 
CaF$_2$ detectors immersed in a $4 \pi$ active shield made out of liquid scintillator. \\
The CANDLES~III experiment is presently taking data since Spring 2013 at the Kamioka 
underground laboratory, Japan. It is operating 96 CaF$_2$ detectors 
(305~kg of total mass, but with natural $^{48}$Ca abundance) immersed in liquid 
scintillator~\cite{candles3}. Due to the usage of non-enriched Ca, CANDLES~III cannot 
be competitive in terms of sensitivity to \onbb\ decay, but the experimental performance 
(e.g. energy resolution) and the achievable background are being studied and optimized. 
The CANDLES~III set up is scheduled to take data for 2014 and 2015, before the transition to 
the next upgrades (CANDLES~IV and CANDLES~V), in which detectors enriched 
in $^{48}$Ca will be deployed, provided that the current R\&D on the enrichment is successfully 
completed. However, the funding for the next phases of CANDLES has not been secured yet.

\section{The \gerda\ experiment}
The \gerda\ experiment at the Gran Sasso National Laboratory (LNGS) of INFN, 
Italy, is searching for the \onbb\ decay of \gess~\cite{gerda:2013:tec}. 
The recent physics results from the Phase~I data~\cite{gerda:2013:prl}
are summarized and reviewed here. 
\subsection{The experimental set up}
The operational design of the \gerda\ experiment 
follows the concept proposed in Ref.~\cite{heus95}: high-purity 
germanium (HPGe) detectors made out of material isotopically enriched in 
\gess\ (\geenr) are operated naked in liquid argon. The liquid argon 
is contained in a stainless steel cryostat of 4~m radius: it provides at the 
same time a very radio-pure passive shielding against the external radiation 
and the cooling which is necessary to operate HPGe detectors. The cryostat 
is enclosed in a 3~m-thick water volume, which gives additional shielding 
against external $\gamma$-rays and neutrons. The water is equipped with 
66~photo-multipliers and is operated as a Cherenkov veto, to identify events 
in the detectors that are originated by muon-induced showers. The \geenr\ 
detectors are deployed within vertical strings containing two or three 
elements each. Neighbor strings are kept very close to maximize the 
background rejection by anti-coincidence in the detector array: genuine \onbb\ 
decays mostly release the total amount of energy (\qbb=2039~keV) in one 
detector only. Particular care was devoted to reduce the amount of material close 
to the detectors (holders, cables, electronics, etc.), and to maximize the 
radio-purity of all components. \\
In the first Phase of \gerda, which lasted from November 2011 to May 2013, the 
existing coaxial \geenr\ detectors previously operated by the 
\hdm~\cite{hdm} and \IGEX~\cite{igex,igex2} experiments were re-used. The data 
from six out of the eight detectors could be considered for the physics analysis (14.6~kg); 
the other two detectors exhibited high leakage current or other 
instabilities. In June 
2012, five newly produced \geenr\ detectors of BEGe type (manufactured 
by Canberra) 
were deployed~\cite{gerda:2013:tec}. They are the outcome of the first 
production 
batch of \gerda\ custom-made \geenr\ detectors, after their initial test-bench 
characterization~\cite{depbege2013,heroica2013}. The data of four out of the 
five BEGe 
detectors (3.0~kg) could be used for physics analysis. \\
The total exposure accumulated 
within the \gerda\ Phase~I is 21.6 \kgyr, with a duty factor of about 
88\%~\cite{gerda:2013:prl}. 
A temporary increase of the background 
of the coaxial detectors in the energy range of interest for the \onbb\ searches 
was observed 
after the deployment of the BEGe detectors; the background came back to the 
previous 
level in approximately 20~days~\cite{gerda:2013:bck}. The data of the coaxial 
detectors 
taken during this period was labeled as ``silver data set'' 
(1.3~\kgyr\ exposure), while all the rest was labeled as 
``golden data set'' (17.9~\kgyr). The BEGe data available for physics 
analysis amount to 2.4~\kgyr. \\
The energy reconstruction is 
performed off-line by using a semi-gaussian filter~\cite{gelatio}. The 
exposure-averaged 
energy resolution at \qbb\ is $(4.8\pm0.2)$~keV and $(3.2 \pm 0.2)$~keV 
full width at 
half-maximum (FWHM), for the coaxial and BEGe detectors, 
respectively~\cite{gerda:2013:bck}. 
The energy scale was determined and monitored by one-hour irradiations with 
three $^{228}$Th radioactive sources, that were performed every one or two 
weeks. 
The shift of the position of the 2615~keV $\gamma$-line from the $^{208}$Tl 
decay between two consecutive calibrations is about 0.5~keV rms, which is much 
smaller than the characteristic energy resolution of the detectors. 
In addition, the stability of the electronics was monitored in real-time 
by regularly injecting charge pulses of fixed amplitude into the input 
of the amplifiers.
\subsection{Data analysis and results}
\begin{figure}[tb] 
\includegraphics[width=0.9\columnwidth]{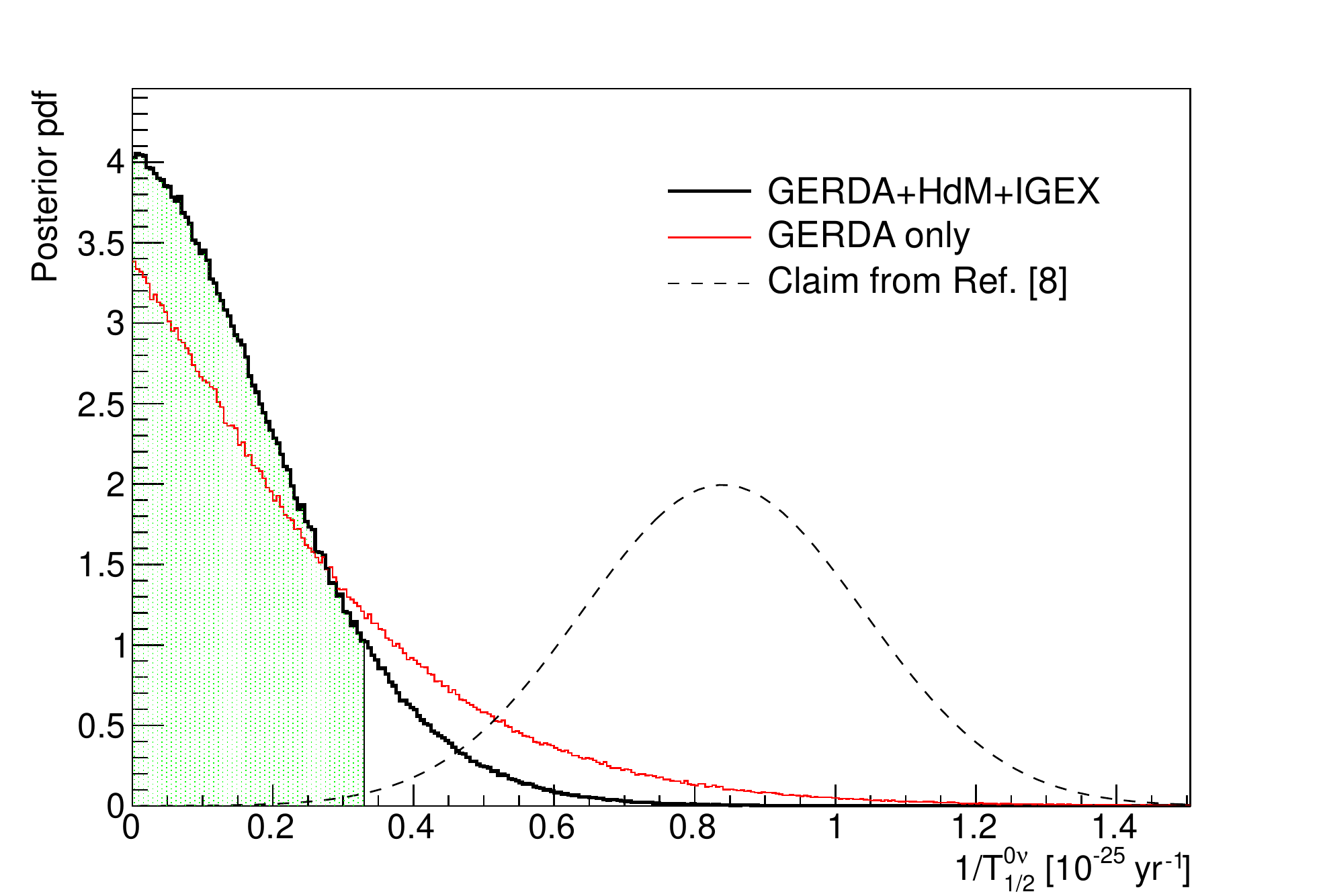} 
\caption{\label{fig:posterior} Posterior pdf $p$(1/\thalfzero) for the inverse half-life 
derived from the Bayesian analysis. The red (thin) and black (thick) solid lines are the 
posterior distributions 
obtained for the \gerda\ data alone and for the combination of \gerda\ with \IGEX\ and \hdm, 
respectively. The dashed peak shows the observation claim according to Ref.~\cite{klapdor1}. 
The shaded area in green covers the 90\% credibility interval for the combined analysis 
(1/\thalfzero $< 0.34 \cdot 10^{-25}$~yr$^{-1}$).}  
\end{figure}

Two main paradigms drove the analysis strategy of \gerda\ Phase~I:
\begin{itemize}
\item blind analysis. All events in a 40~keV range around \qbb\ were initially 
not made available for the analysis (neither the energy nor the pulse shape). 
Background model and pulse shape discrimination techniques were 
developed and validated on the open data. All procedures and cuts were 
frozen before the unblinding. 
\item all valid physics data are considered for the analysis. The available 
data are separated in three data sets, which differ for background and 
energy resolution, and a combined analysis is performed.
This approach allows to maximize the amount of data which is accounted for 
the analysis, and avoids the lower-quality data (e.g. the 
``silver'' data set) to spoil the sensitivity of the higher-quality ones, because 
data are never summed up. 
\end{itemize} 
A complete and quantitative background model was developed and validated 
before the unblinding, using a fraction (85\%) of the final data 
set~\cite{gerda:2013:bck}. The predictions provided by the model were 
checked ``a posteriori'' against the final data set (without any 
additional fit) and were found to be consistent with the events 
uncovered from the blinded region.
The main outcomes from the background model  
affecting the \onbb\ analysis were that: (1) the expected background at \qbb\ has 
a flat energy spectrum in a relatively large range; (2) no intense 
$\gamma$-lines are expected in the vicinity of \qbb. As a consequence, the 
experimental \gerda\ spectrum was fitted by using a flat background model 
between 1930 and 2190~keV (apart from two known $\gamma$-lines at 2104~keV and 
2119~keV). \\
\begin{figure}[tb] 
\includegraphics[width=0.9\columnwidth]{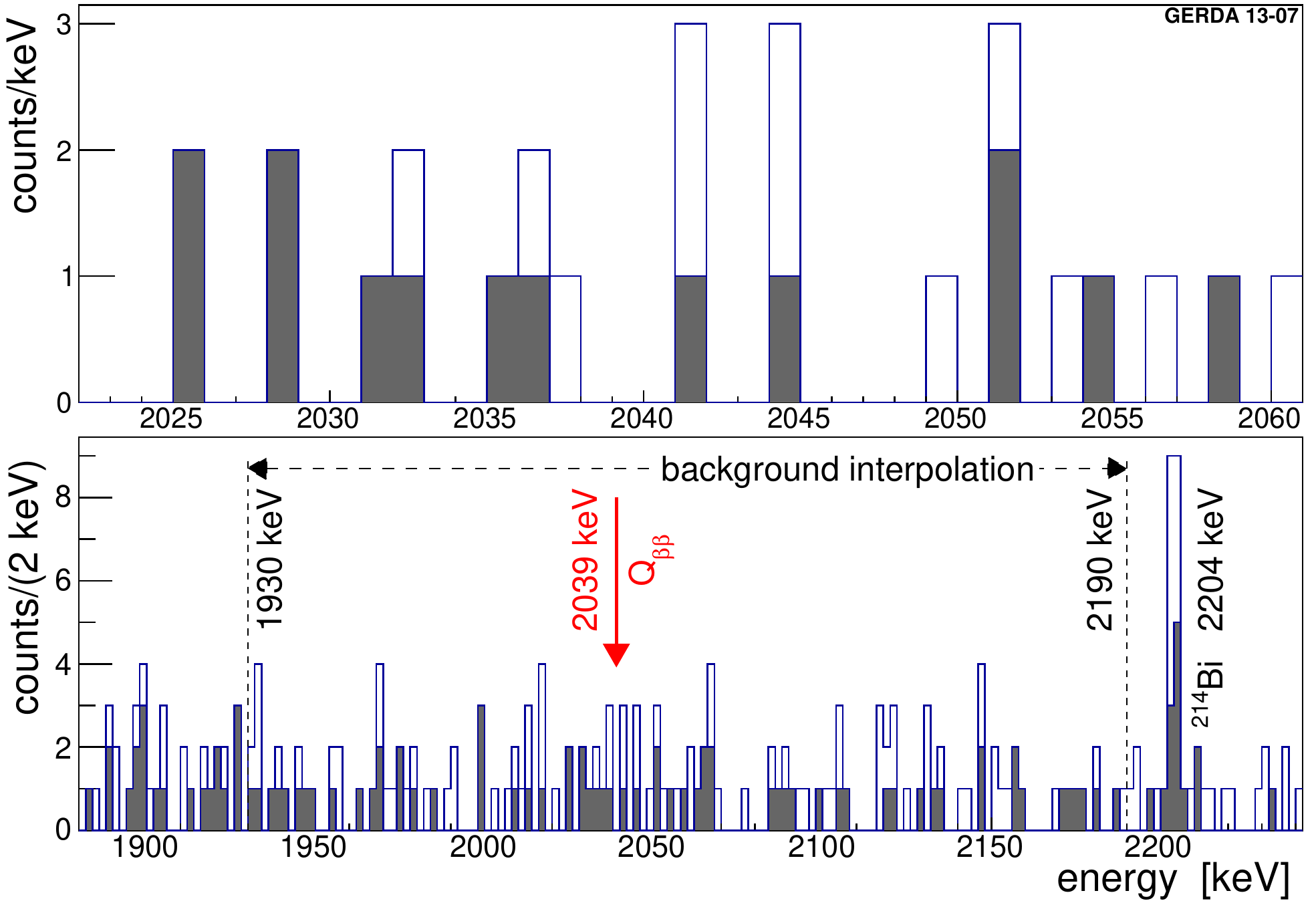} 
\caption{\label{fig:spectrum}  Sum energy spectrum of \gerda\ in the \qbb\ region, 
before (open histogram) and after (filled histogram) pulse shape discrimination. 
The energy region used 
for the background interpolation is shown in the lower panel. The only $\gamma$-line 
which is positively identified in the energy range of the lower panel is the one at 
2204~keV from $^{214}$Bi.}  
\end{figure}
The sum energy spectrum of \gerda\ is displayed in Fig.~\ref{fig:spectrum}, 
with and without the pulse shape discrimination (PSD) described in 
Ref.~\cite{gerda:2013:psd}. Notice that the spectra of the three data sets 
are shown here together, but they were considered separately for the  
analysis. The typical background level achieved in the coaxial detectors before PSD 
was about $2 \cdot 10^{-2}$~\ctsper. The effect of the PSD was to reduce the continuous 
background by approximately a factor of two for the coaxial detectors and 
by a factor $> 5$ for the BEGe detectors, for a \onbb\ signal acceptance of 
90\% and 92\%, respectively. Since the sensitivity of the experiment scales 
as signal/$\sqrt{\textrm{background}}$, the optimal configuration 
for the \gerda\ PSD cuts was achieved by keeping a high acceptance and 
a moderate background rejection~\cite{bern}. 
More strict PSD cuts $-$ which further reduce the residual background at the price 
of a lower signal acceptance $-$ do not provide any improvement in sensitivity. \\
Having fixed all cuts and procedures in advance, the data unblinding revealed 
seven events in the energy range 
$Q_{\beta\beta} \pm 5$~keV, as summarized in Table~\ref{tab:events}; three 
events survived the PSD analysis, at energies 2036.9~keV (``silver''), 
2041.3~keV (``golden'') and 2035.5~keV (``golden''), respectively. No events 
were found within $\pm 1 \sigma$ from the \qbb\ value, being $\sigma$ 
the expected (root mean square) width of the Gaussian peak due to energy 
resolution. The number of events at \qbb\ is consistent with the 
expectations from a constant background, as derived from the energy 
range 1930--2190~keV (see Table~\ref{tab:events}). \\
There is no indication of  
unidentified $\gamma$-lines in the region of interest \qbb$\pm 20$~keV.
In particular, the possible presence of 
weak $\gamma$-like structures at 2016~keV and 2052~keV 
from the $^{214}$Bi decay\footnote{The structure at 2016~keV would emerge 
as a combination of a line at 2016.7~keV and two unresolved lines at 
2010.8 and 2021.6~keV, respectively.} can be assessed quantitatively by 
the comparison with the more intense $^{214}$Bi line at 
2204~keV~\cite{gerda:rev:macolino}. 
The lines at 2016~keV and 2052~keV were 
observed in the \hdm\ spectrum~\cite{klapdor2001}: they were a major 
discussion topic in the early times after the publication of the 
observation claim~\cite{klapdor1,aals2002,kla02,bilines}. The intensity of the 
2204~keV line (5.08\% branching ratio) in \gerda\ is 
($0.8 \pm 0.3$)~cts/\kgyr~\cite{gerda:2013:bck}, before PSD, 
corresponding to $17.3 \pm 6.5$~counts. This is an order of magnitude lower 
than the level observed in  \hdm, ($8.1 \pm 0.5$)~cts/\kgyr~\cite{olegphd}. 
The number of counts expected in the \gerda\ data for the structures at 2016~keV 
and 2052~keV is $< 1$, as summarized in Table~\ref{tab:bi}. \\
\begin{table}[tbhp] 
\centering \caption{\label{tab:events} Summary of the events detected by \gerda\ in the 
energy range \qbb$ \pm 5$~keV, before and after the PSD cuts. The events are subdivided 
in the three reference data sets (``golden coaxial'', ``silver coaxial'' and BEGe). The 
expected number of background events is also shown, as derived 
from the interpolation range of Fig.~\ref{fig:spectrum}.}
\begin{tabular}{l|cc|cc} 
Data set & \multicolumn{2}{|c|}{Without PSD} & \multicolumn{2}{|c}{With PSD}\\
 & observed & expected & observed & expected \\
\hline
golden & 5 & 3.3 & 2 & 2.0 \\
silver & 1 & 0.8 & 1 & 0.4 \\
BEGe & 1 & 1.0 & 0 & 0.1 \\
\end{tabular} 
\end{table}
\begin{table}[tbhp] 
\centering \caption{\label{tab:bi} Evaluation of the intensity of the weak $^{214}$Bi lines at \qbb\ 
in \gerda\ from the $^{214}$Bi line observed at 2204~keV. The full energy peak efficiency is assumed to 
be approximately the same at 2016, 2053 and 2204~keV, so intensities are scaled according to the 
branching ratios only.}
\begin{tabular}{lcc} 
Line (keV) & Branching ratio & Intensity (counts) \\
\hline
2204.2 & 5.08\% & $17.3 \pm 6.5$ \\
\hline
2010.8 + 2016.7 + 2021.6 & 0.067\% & (0.23) \\
2052.9 & 0.069\% & (0.23) \\
\hline
\end{tabular} 
\end{table}
Calibration data taken with $^{228}$Th and $^{56}$Co sources proved that
all $\gamma$-ray peaks and 
all double-escape events (that are kinematically mimicking the \onbb\  
decay) are reconstructed at the correct position~\cite{bern}. This confirms 
that there are no significant effect of ballistic deficit in \gerda, and 
that the \onbb\ signal is expected at \qbb. \\
The limit for the number of events in the \onbb\ peak at \qbb\ was  
derived by a combined maximum likelihood fit of the energy spectra. The 
three data sets were analyzed individually, each with its own average 
background (free parameter) and energy resolution. The inverse half-life of 
the \onbb\ decay 1/\thalfzero (which is proportional to the number of counts) 
is kept as a common free parameter. The analysis was performed with a 
frequentist (baseline) and a Bayesian approach, using the same likelihood 
function~\cite{gerda:2013:prl,bern}. The best fit is obtained for 1/\thalfzero=0, 
namely, no counts ascribed to the \onbb\ decay. The frequentist analysis gives a limit 
$T_{1/2}^{0\nu} > 2.1\cdot10^{25}$~yr, at 90\% C.L., to be compared with a 
median expected sensitivity of $2.4\cdot10^{25}$~yr. When the data from the 
predecessor experiments \hdm~\cite{hdm} and \IGEX~\cite{igex} are included into a 
combined fit, the limit is strengthened to $T_{1/2}^{0\nu} > 3.0\cdot10^{25}$~yr, 
at 90\% C.L. The likelihood profiles obtained in this analysis are shown in 
Ref.~\cite{bern}. \\ 
For the Bayesian approach, a flat prior probability distribution was taken for 
1/\thalfzero between 0 and $10^{-24}$ yr$^{-1}$.The marginalized posterior distribution 
$p$(1/\thalfzero) is shown in Fig.~\ref{fig:posterior}, for the \gerda\ data and 
for the combination with 
\IGEX\ and \hdm. The most probable value is 1/\thalfzero=0 in both cases. The 90\% probability 
quantile of the \gerda\ posterior distribution is $T_{1/2}^{0\nu} > 1.9\cdot10^{25}$~yr 
(90\% credible interval), for a median sensitivity of $2.0\cdot10^{25}$~yr. The limit 
coming from the combined analysis with \hdm\ and \IGEX\ is 
$T_{1/2}^{0\nu} > 2.9\cdot10^{25}$~yr (90\% credible interval). \\

The negative result obtained by \gerda\ (also in combination with the predecessor 
experiments) can be compared quantitatively against the observation claim of 
Ref.~\cite{klapdor1}. The result given in Ref.~\cite{klapdor2} is not considered 
here because of the inconsistencies pointed out in Ref.~\cite{bernhard}. 
The hypothesis $H_1$, which is a \onbb\ decay with 
\thalfzero$=(1.19^{+0.37}_{-0.23}) \cdot 10^{25}$~yr, is compared to the null 
hypothesis $H_0$, which is background only. The hypothesis $H_1$ predicts 
$(5.9 \pm 1.4)$ events from the \onbb\ decay in $Q_{\beta\beta} \pm 2 \sigma$ sitting on a 
constant background of $(2.0 \pm 0.3)$ events, after all PSD cuts. There are actually 
3 counts observed by \gerda\ in $Q_{\beta\beta} \pm 2 \sigma$, and none of them is in 
$Q_{\beta\beta} \pm 1 \sigma$. \\
A set of $10^4$ Monte Carlo repetitions of the \gerda\ experiments was generated with 
the assumption of the hypothesis $H_1$, to evaluate in which fraction of the cases the 
profile likelihood analysis yields 1/\thalfzero=0 as the best fit~\cite{bern}. It was found 
that P(1/\thalfzero=0 $| H_1$)=0.01, i.e. the probability to produce the actual outcome 
of \gerda\ with the \onbb\ signal reported in Ref.~\cite{klapdor1} is 1\%. From the 
Bayesian analysis it was also possible to derive the Bayes factor, i.e. the odd ratio 
between the two hypotheses under testing. It is P(H$_1$)/P(H$_0$)=0.024 with the \gerda\ 
data alone, which is further reduced to P(H$_1$)/P(H$_0$)=0.0002 when the \IGEX\ and 
\hdm\ data are included in the fit. The long-standing claim for a \onbb\ signal in \gess\ 
is hence strongly disfavored. 
\subsection{Transition to Phase~II}
The transition to the Phase~II of \gerda\ is presently ongoing. The goal of the Phase~II 
is to increase by an order of magnitude the sensitivity on the \thalfzero\ of \gess, 
namely to the scale of a few $10^{26}$~yr~\cite{belataup}. 
This will be achieved by the increase 
of the total detector mass and by the further suppression of the background at \qbb, 
down to $10^{-3}$~\ctsper. About 30 custom-made \geenr\ BEGe detectors are available 
for \gerda\ Phase~II, totaling about 20~kg mass. They have been produced by Canberra 
Olen and they have been accurately characterized at the HADES underground 
facility, Belgium~\cite{heroica2013}. The first batch of the production (5 \geenr\ BEGe 
detectors) was deployed in \gerda\ Phase~I and hence tested in the real-life 
environment. No anomalies with internal or surface contamination were 
observed in the newly produced detectors; in particular, the surface contamination 
from $^{210}$Po turned out to be much smaller than for the coaxial detectors.\\ 
There are two main handles to achieve a background reduction by one order 
of magnitude with 
respect to Phase~I: (1) further reduce the amount and the radioactivity of the 
materials in close vicinity of the detectors; (2) better reject the residual 
background by the powerful PSD of the BEGe 
detectors~\cite{aovere} and by the 
instrumentation of the liquid argon volume surrounding the detector array as an 
active veto. \\
A new front-end read-out and cabling are being developed with better 
radio-purity: this allows to place the front-end very close to the detectors ($< 2$~cm) 
and hence to improve the energy resolution. Furthermore, new high voltage and 
signal cables are made available, with improved radio-purity and $^{222}$Rn 
emanation. The scintillation light of liquid argon in the 500~liter volume surrounding 
the HPGe 
detector array will be detected by photo-multipliers and SiPM detectors, thus turning 
the volume into an active veto system, which is very effective for the identification 
of background coming from external $\gamma$ sources. \\
The commissioning of the Phase~II set up is expected to start before Summer 2014. 
\section{Conclusions}
Studies of the most popular candidate isotopes for the \onbb\ decay indicate that there is not a 
theoretical ``golden isotope'', which is clearly favored in terms of specific 
decay rate for a given Majorana neutrino mass. In spite of the differences in the 
nuclear matrix elements and in the phase space factors, all isotopes yield approximately 
the same decay rate. Therefore, other technical and 
practical parameters enter into the play $-$ as energy resolution, background, scalability 
and cost $-$ which can easily compensate for a less favorable nuclear matrix elements 
or a lower \qbb-value. \\
There are presently many experimental and R\&D programs ongoing, taking different candidate 
isotopes into consideration, that aim to reach an exposure in the scale of hundreds of 
\kgyr\ within the next 5-10 years. In particular: (1) \gerda\ has completed the data taking 
for Phase~I (21.6~\kgyr) and is currently upgrading the set up for the Phase~II; (2) 
Cuore-0 recently started the data taking, as the first step of the ton-scale Cuore 
project; (3) other experiments are being built, commissioned or designed. New results about 
the \onbb\ decay of $^{100}$Mo have been recently released by NEMO3.\\
The data collected in \gerda\ Phase~I were subject to a blind analysis. No positive 
signal was found at the \qbb-value of \gess, with a background of approximately 
$10^{-2}$~\ctsper, after the pulse shape discrimination. A lower limit on the \onbb\ 
decay is set to \thalfzero $> 2.1 \cdot 10^{25}$~yr at 90\% C.L. (\gerda\ alone), or 
\thalfzero $> 3.0 \cdot 10^{25}$~yr (in combination with the predecessor experiments 
\hdm\ and \IGEX). The observation claim of Ref.~\cite{klapdor1} is hence strongly 
disfavored by the \gerda\ data. The comparison is model-independent, since it is 
referred to the half-life \thalfzero\ of \gess\ and does not involve theoretical 
calculations of nuclear matrix elements. The next phase of \gerda, which will start 
within the next few months, aims to collect an exposure of 100~\kgyr\ in 3 years, with 
a further reduced background, to push the sensitivity on \thalfzero\ to the $10^{26}$~yr 
range.


\section*{Acknowledgments} 
This paper is the write up of a talk given at the XIII
International Conference on Topics in Astroparticle and Underground
Physics (TAUP2013), in Asilomar, CA, United States (8-13 September,  
2013). The author would like to thank again the TAUP2013 organizers, 
and especially the co-chairs of the Organizing Committee, W.~Haxton 
and F.~Avignone, for the kind invitation. Material and suggestions 
for the original talk were provided by A.~Bettini, M.~Bongrand, 
M.~Chen, O.~Cremonesi, 
S.~Elliott, F.~Piquemal and S.~Umehara, who are kindly acknowledged. The 
author thanks the \gerda\ Collaboration for the designation and 
for many useful comments and feedback to the talk and to this paper.

\bibliographystyle{elsarticle-num} 
\bibliography{biblio}




\end{document}